Beyond Galileo:

A translation of Giovanni Battista Riccioli's experiments regarding falling bodies and "air drag", as reported in his 1651 *Almagestum Novum*

Christopher M. Graney

Jefferson Community & Technical College

Louisville, Kentucky (USA), 40272

christopher.graney@kctcs.edu

The Italian astronomer Giovanni Battista Riccioli is commonly credited with performing the first precise experiments to determine the acceleration of a freely falling body, but he also went further, experimentally investigating what today would be called the effect of "air drag" on falling bodies. This paper consists of a translation of those experiments, with a brief analysis and commentary. Riccioli arrived at conclusions consistent with modern understanding of "air drag".

The 1651 *Almagestum Novum* by Giovanni Battista Riccioli (1598-1671), an Italian astronomer, was an influential, massive, encyclopedic[1] treatise discussing a wide variety of subjects. One of these was the behavior of falling bodies, following up on Galileo's work in this area. Indeed, Riccioli's falling body experiments include what are often cited as the first precise experiments to determine the acceleration due to gravity (Koyré 1953, 231-232; Koyré 1955, 349; Lindberg & Numbers 1986, 155; Heilbron 1999, 180). A translation of those is now available (Graney 2012a). But Riccioli also investigated what today might be termed the effect of "air drag" on falling bodies, by dropping balls of differing sizes and weights from a fixed height, and comparing their relative rates of fall.

Riccioli achieved excellent results in his "acceleration" experiments, and as will be seen below, his "air drag" experiments also yielded excellent results. Full translations of Riccioli's work from Latin into modern languages are not readily available. Thus a translation of Riccioli's experiments is of value.

The text translated here is heading IV in Chapter XXVI of Book IX, Section IV of the second volume of the *Almagestum Novum*, pages 387-9. These headings and pages contain the "air drag" experiments with falling bodies. They directly precede further experiments on bodies falling through water. Included in footnotes and following the translation is some brief commentary on Riccioli's work and some discussion of his conclusions seen in the light of modern ideas.

We (I thank Christina Graney for her invaluable assistance in translating Riccioli's Latin) intend this translation to hew closely to Riccioli's original work. However, Riccioli uses many different sorts of verbs, sentences that run in excess of 100 words in length, and few paragraphs, among other things. Therefore, we use simpler verbs. We turn long sentences into multiple short sentences. We break long paragraphs into short ones (a double-space indicates Riccioli's original paragraphs; indents indicate where we add a paragraph break). We retain Riccioli's numbering system, which uses one set of Roman numerals for paragraphs and another for headings. Places where the translation departs significantly from the original are noted as such.

---

[1] The *Almagestum Novum* consists of over 1500 large-format pages, mostly of dense type with some diagrams.

*IV. The Group of Experiments about Unequal Descent of Two Heavy Bodies of unlike weight through the Air from the same altitude.[2]*

XIII. This experiment is very misleading unless it is completed with great diligence and circumspection. All present must be able to reach the same conclusions concerning which motions will be faster or slower.

The best arrangement, which would remove all doubts, would be if two balls of the same material and same appearance but different weights would be released from the same height at the same time. Imagine that we could take two clay or metal balls — one a palm's width in diameter and weighing one Roman pound, the other two palms and eight pounds — and rarefy the smaller one while retaining its solidity (without adding any foreign material to it) so that it became the size of the larger: Then if in fact the heavier was found to descend more quickly, the quicker descent could be attributed only to the strength of greater heaviness, and not to difference of appearance or bulk, nor to difference of shape (where a larger ball would have to break through the broader expanse of air below it). In short, the best arrangement is one of parity in the case of all things except weight.

It is truly not expedient to release two cylinders or two prisms made from the same material, of equal width, but of unequal height or weight. They fluctuate around their centers of gravity in the air, and they leave behind uncertain cause of a swifter or tardier descent.

The best arrangement is not possible, and so Fr. Grimaldi[3] and I prepared what is nearest to it: twelve clay balls of solid clay, of which the weight of each was 20 ounces; and twelve others from clay, equal in bulk to the 20 ounce balls, but hollow on the inside, each weighing 10 ounces.

---

[2] See Graney 2012a for Riccioli's details about how falling body experiments were conducted.
[3] Francesco Maria Grimaldi (1618-1663), Riccioli's assistant.

Then we acquired many other diverse balls, differing in bulk, or in weight. We then prepared different comparisons.

In May 1640, August 1645, October 1648, and most recently in 1650 we released different pairs from the crown of the tower of Asinelli, before many witnesses convoked from our Society.[4] Although they have each not always been to the same experiment, these include: *Frs. Stephanus Ghisonus, Camillus Rodengus, Iacobus Maria Pallauicimus, Franciscus Maria Grimaldus, Vincentius Maria Grimaldus, Franciscus Zenus,* and *Paulus Casatus* with his students *Franciscus Adurnus* and *Octavius Rubeus.* Each is distinguished by character, judgment, and religious integrity. And indeed among these, three or four Masters of Philosophy or Theology were present, who with Galileo, Cabeo, or Arriaga, had judged that any two heavy bodies, released simultaneously from the same altitude, however great, descend to the ground by the same natural moment of time.

But they promptly gave up this opinion, since a lighter (10 ounce) clay ball — released from G [see Figure 1] at the same moment as a second clay ball of the same bulk but of 20 ounces was released from O — was seen to be at F, distant from the pavement I by at least 15 feet, at the same moment at which the heavier ball struck the same pavement at D and burst into myriad fragments. Although before the arrival of each to the pavement the lighter ball (which by agreement Fr. Grimaldi always released from the left hand) would appear greatly separated from the heavier one a little below the middle of the tower. In terms of time, counted by the help of the pendulum, the heavier ball descended by exactly 23 strokes, or 3 and 1/2 seconds, the lighter by 26 strokes, or 4 and 1/3 seconds.[5] This experiment was repeated twelve times and timing pendulum, 23 strokes is 3 and 5/6 seconds rather than the 3 and 1/2 seconds stated here. Also, the 15 foot distance gap between the heavier and lighter balls agrees reasonably well with modern calculations but the 3 stroke time gap does not (simulations using reasonable estimates for the sizes and densities of the balls Riccioli dropped — "palm-wide" being roughly 9 cm, for example — produce a gap of very roughly 15 feet relatively easily, but a 3 stroke time gap cannot be reasonably simulated). Finally, the 3 stroke time gap does not agree with Riccioli's writings a few pages earlier in the *Almagestum Novum*, where he concluded that a falling clay ball of 8 ounces takes 5 strokes to fall 10 Roman feet from rest (and 26 strokes to fall 280 Roman feet), and that velocity increases linearly with time. Thus he says, "during the first measure it traverses OC, 10 feet; during the second it traverses CQ, 30 feet; third, QR, 50 feet; fourth, RS, 70 feet; and fifth, ST, 90 feet [Graney 2012a, 14]". For the 10 ounce ball to travel just 15 feet in 3 strokes at around the 4th second of fall is inconsistent with these numbers.

---

[4] "Our Society" is the Society of Jesus — Riccioli was a Jesuit.

[5] The 23 stokes value is probably a typographical or transcription error. One second was 6 strokes of Riccioli's

**Figure 1**

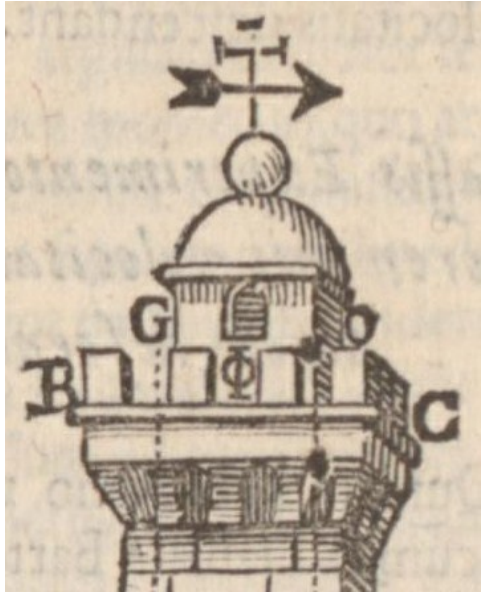
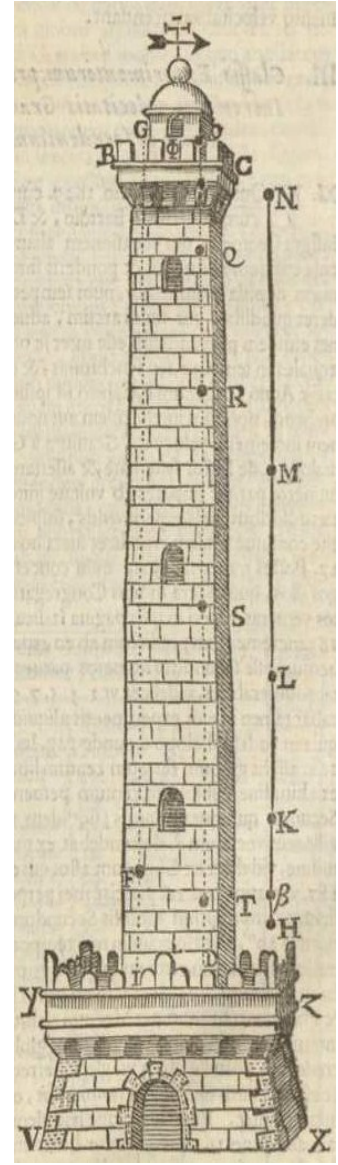
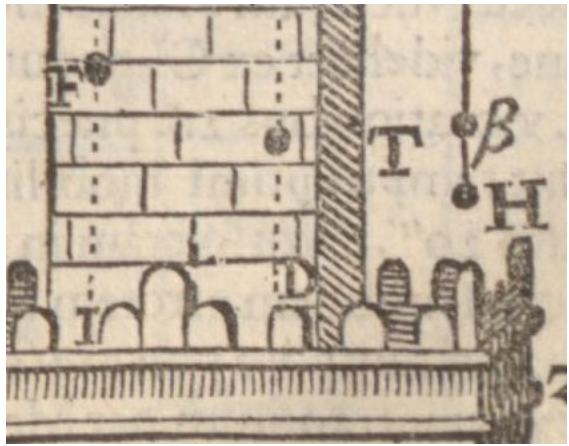

# Riccioli's Data Table

**Balls released at the same moment from an altitude of 280 Feet**

| Order of the Experiments | Balls of the same size | Weight of the balls | | Ball which was faster | Distance of the slower from the pavement when the faster struck (feet) |
|---|---|---|---|---|---|
| | | **Ounces** | **Drachmas** | | |
| 1 | Clay hollow | 10 | 0 | | |
|   | Clay solid  | 20 | 0 | Heavier | 15 |
| 2 | Clay hollow | 20 | 0 | | |
|   | Clay solid  | 45 | 0 | Heavier | 16 |
| 3 | Clay | 9 | 0 | | |
|   | Wood | 2 | 4 | Heavier | 20 |
| 4 | Clay | 20 | 0 | | |
|   | Wax  | 15 | 0 | Heavier | 12 |
| 5 | Wood | 4 | 6 | | |
|   | Wax  | 6 | 7 | Heavier | 15 |
| 6 | Wax  | 1  | 5 | | |
|   | Iron | 11 | 7 | Heavier | 30 |

Experiment 1 was repeated 12 times, experiment 2 was repeated twice.

**Second part of the Preceding Table**

| Order | Balls of unequal size | Weight of the balls | | Ball which was faster | Distance etc. |
|---|---|---|---|---|---|
| | | Ounces | Drachmas | | |
| 7 | Clay | 5 | 0 | | |
| | Clay | 4 | 0 | Heavier | 5 |
| 8 | Clay | 21 | 4 | | |
| | Clay | 11 | 4 | Heavier | 12 |
| 9 | Clay | 27 | 3 | | |
| | Clay | 14 | 1 | Heavier | 15 |
| 10 | Clay | 18 | 7 | | |
| | Wax | 1 | 0 | Heavier | 35 |
| 11 | Clay | 2 | 0 | | |
| | Lead | 2 | 4 | Heavier | 25 |
| 12 | Lead | 2 | 4 | | |
| | Wood | 2 | 4 | Lead | 40 |
| 13 | Clay | 7 | 0 | | |
| | Clay | 62 | 0 | Heavier | 10 |
| 14 | Clay | 10 | 4 | | |
| | Clay | 23 | 0 | Heavier | 13 |
| 15 | Clay | 53 | 0 | | |
| | Clay | 6 | 4 | Heavier | 8 |
| 16 | Clay | 53 | 0 | | |
| | Clay | 7 | 1 | Heavier | 9 |
| 17 | Walnut Wood | 2 | 1 | | |
| | Beech Wood | 4 | 7 | Walnut | 2 |
| 18 | Clay | 11 | 0 | | |
| | Lead | 1 | 7 | Lead | 1 |
| 19 | Clay | 33 | 0 | | |
| | Lead | 1 | 0 | Clay | 2 |
| 20 | Clay | 38 | 0 | | |
| | Lead | 1 | 0 | Clay | 3 |
| 21 | Lead | 1 | 0 | | |
| | Lead | 0 | 4 | Heavier | 5 |

always had a similar outcome.

Since the distinction between the two balls was more easily observed in the difference of distance than in the difference of time, the decision was to attend to the residual distance interval of the more slowly descending ball FI, than to the time interval, as seen in the following table (in which the earlier experiments consider two balls of the same bulk, while the latter experiments consider balls of different bulk).

*Conclusions from the preceding Table.*

XIV.  Here I suppose a one body to have greater specific Heaviness than another if, bulk being equal, it is greater in weight, as lead is said to be heavier than wax, because if two palm-wide balls may be weighed, one lead, the other wax, the lead will weigh either more pounds or ounces or drachmas etc. than the wax.  By contrast, a body has greater individual Heaviness, whether it is of the same specific heaviness or not, if it is of greater weight absolutely, just as a wax ball of a hundred pounds is said to be heavier than a lead ball of one ounce.  Hence five useful combinations arise:[6]

1. Those where both objects are equally heavy specifically and individually — for example, two lead spheres, of one pound each.  These are naturally of the same bulk.  If one were rarified supernaturally or preternaturally, it could reach a greater bulk.
2. Those where both objects are equally heavy specifically, but not individually — for example two lead balls, where one weighs one, the other two pounds.  If these both be solid, then that which is the heavier individually — the two pound ball — is also the larger; by supernatural or preternatural compaction it might be made to be equal to or lesser than the other. And the lighter ball might equal the heavier one in bulk if made hollow on the inside, or full of air.
3. Those where both objects are equally heavy individually, but not specifically — for example, a lead ball and a wax ball, each of which is one pound. Naturally that which is the lighter specifically (the wax) is the larger; if it could be compacted supernaturally or preternaturally it could be made to be no bigger than the other.
4. Those where one object is Heavier both specifically and individually — for example a 10 pound lead ball compared to a one pound wax ball. The heavier object can be of equal bulk, or lesser, or greater.
5. Those where one object is heavier specifically, but lighter individually — for example, a little lead ball of one ounce and a wax ball of one hundred pounds. In such a case the first is naturally smaller than the second. Here the lead ball is heavier specifically, while the wax ball is heavier individually.

---

[6] Here this translation deviates significantly from Riccioli's original Latin.  Riccioli wrote the five combinations out in paragraph form, and not in a numbered list.

Granted these, the conclusions below follow from the preceding table.

*First:* Two spheres, equally heavy both specifically and individually, naturally descend equally rapidly through the same medium (if released simultaneously from the same altitude). They are of equal bulk, so all things are like. Yet suppose one of those might be made (supernaturally or preternaturally) either greater by rarefaction, or lesser by compaction: the smaller sphere would descend faster; the contact angle, by which it might hit the plane of the underlying air, would be more acute.

*Second:* Of two spheres equally heavy specifically but not individually, the one which is heavier individually naturally descends more quickly through the air. This is true whether they are of equal bulk (because one of them is hollow within), as in experiment 1 (repeated twelve times) and experiment 2; or of differing bulk, as in experiments 7, 8, 9, 13, 14, 15, 16, and 21. Some of these I believe may indicate unknown proportional relationships between speed and the individual heaviness, the size, and the lack of roundness of a ball. However, it is better to report experiments faithfully rather than touch them up by selecting out certain results.

*Third:* Of two spheres equally heavy individually, but not specifically, the one which is heavier specifically naturally descends more quickly through air, as seen in experiment 12. The reason for this is that the one which is heavier specifically is naturally of smaller bulk than the other. Thus the one that is heavier specifically is also sharper of shape or angle of contact (owing to the smallness of the sphere). This is apart from the fact that in our case, the wood sphere is more porous than the lead, and less suitable to gravitating, on account of the levity[7] of the air or vapors hidden within the pores.

*Fourth:* Of two spheres, one of which is heavier not only specifically, but also individually, the heavier one naturally descends more swiftly through the air, whether it is larger than the other, as in experiment 10, or equal to the other, as in experiments 3, 4, 5, 6, or indeed smaller, as in experiment 11.

---

[7] Note that Riccioli is writing well before the advent of Newtonian physics. A geocentrist who mustered strong scientific arguments against the heliocentric theory (Graney 2012b), he understands different materials to have the intrinsic properties of being heavy, or having "gravity" (and thus seeking to reach the center of the universe, which is the center of the Earth), or of being light, or having "levity" (and thus seeking to rise away from the center of the Earth). This is rooted in Aristotelian physics. Aristotle and Riccioli would consider a rock to have gravity and a helium balloon to have levity.

*Fifth:* Of two spheres, of which one is heavier specifically, but not individually, the one which is heavier specifically may descend more quickly, equally quickly, or less quickly through the air. We see an example of the first case in experiments 12, 17, and 18; of the third case in experiments 19 and 20. Moreover we have sufficient argument for the second and third by comparing experiment 8 with 18.

Indeed, in both, the Clay ball was 11 ounces (neglecting drachmas). But in experiment 8 the clay of 11 ounces released with a clay of 21 ounces was slower, being distant 12 feet from the pavement when the 21 ounce clay struck; whereas in experiment 18 the clay of 11 ounces released with the lead of almost 2 ounces was not slower, except by just an interval of one foot. Therefore if the clay of 21 ounces might have been released with the lead of almost 2 ounces, the lead might have trailed it by an interval of 11 feet (or at least by more than one foot). The 21 ounce clay achieved the faster descent.

Therefore with increased individual weight the speed of descent increases to faster from slower. Certainly the individual weight of the 11 ounce clay could be increased a little so as to descend equally swiftly with the lead. *Anyhow* I have said *more than enough*.

Experiments 19 and 20 indeed startled me.[8] In these a little lead ball of one ounce, acquired more (or at least nearly as much) velocity on account of smallness, than the heavier clay acquired on account of far

---

[8] Riccioli expresses surprise to find that air drag can have significant effect. Earlier scientists such as Tycho Brahe (whose work Riccioli often cited) had believed that air was too tenuous to have significant effect on a heavy projectile (see Brahe 1601, 188-92).

greater weight. The more perfect roundness of the lead than of the clay contributed to this.

*Sixth:* Of two balls simultaneously released from the same altitude through the same air, that ball which is lighter in every respect at no time naturally descends more quickly or equally quickly. Indeed, the swifter is either heavier specifically and individually, or heavier individually, or heavier specifically. Hence, said simply, those which are heavier naturally descend more quickly, at least in our air. However, it can happen that two balls are of different weight, and the lack of individual heaviness in one may be compensated for by its sharpness of small bulk, and they descend equally swiftly.

Standard textbook physics tells us that the acceleration $a$ of a spherical body of mass $m$, density $\rho$, radius $r$, cross-sctional area $A$, and drag coefficient $C_d$ falling through air of density $\rho_{air}$ under the influence of its weight ($W = mg$) and air drag $\left(W = \frac{1}{2}\rho_{air}AC_d v^2\right)$ is given by:

$$a = \frac{dv}{dt} = \frac{\Sigma F}{m} = \frac{W - D}{m} = \frac{mg - \frac{1}{2}\rho_{air}AC_d v^2}{m} = g - \frac{AC_d}{2m}\rho_{air}v^2 = g - \frac{\pi r^2 C_d}{2\rho\frac{4}{3}\pi r^3}\rho_{air}v^2$$

$$= g - \frac{3}{8}\frac{C_d}{\rho r}\rho_{air}v^2$$

Thus $a$ is less than $g$ for $v > 0$, and the lesser the density or radius of the ball, and the greater the drag coefficient, the lesser the acceleration at any given speed, and the longer the time required for the ball to drop from a given height.

Riccioli, recognizing that the source of the different fall times of different balls is the interaction of the balls with the air (note he also experimented with bodies falling through water), reaches these same results through his experiments. Riccioli reaches the results explicitly regarding density, or "specific heaviness" [9] — and radius, or size or "bulk" (i.e.

---

[9] The words he uses, "gravius in specie" are really more like "specific gravity", but we did not translate them

volume).  As regards drag coefficient, he simply notes that balls that are less perfectly round are prone to more air drag.

In conclusion, Riccioli's experiment with different falling bodies successfully identified the various factors that determine the effect of air resistance on a falling body. This along with his careful experiments to determine how the speed of a falling body changed over time provided a thorough description of the motion of a falling body — a description published in a prominent and widely-read work.

---

as such because Riccioli's conceptions regarding gravity are not modern.  To Riccioli, gravity is not a force of attraction between Earth and a ball, but a ball's natural tendency to move toward the center of the universe. Thus "heaviness" seemed a more apt translation.